\documentclass[fleqn]{llncs}
\usepackage{isdia}

\title{Instruction Sequences with \\
       Dynamically Instantiated Instructions%
       \thanks{This research was partly carried out in the framework of
               the  Jacquard-project Symbiosis, which is funded by the
               Netherlands Organisation for Scientific Research (NWO).}}
\author{J.A. Bergstra \and C.A. Middelburg}
\institute{Informatics Institute, University of Amsterdam, \\
           Science Park~107, 1098~XG~Amsterdam, the Netherlands \\
           \email{J.A.Bergstra@uva.nl,C.A.Middelburg@uva.nl}}

\begin{document}
\maketitle

\begin{abstract}
We study sequential programs that are instruction sequences with
dynamically instantiated instructions.
We define the meaning of such programs in two different ways.
In either case, we give a translation by which each program with
dynamically instantiated instructions is turned into a program without
them that exhibits on execution the same behaviour by interaction with
some service.
The complexity of the translations differ considerably, whereas the
services concerned are equally simple.
However, the service concerned in the case of the simpler translation is
far more powerful than the service concerned in the other case.
\begin{classcode}
D.3.1, D.3.3, F.1.1, F.3.2, F.3.3.
\end{classcode}
\end{abstract}

\section{Introduction}
\label{sect-intro}

In this paper, we study sequential programs that are instruction
sequences with dynamically instantiated instructions.
With that we carry on the line of research with which a start was made
in~\cite{BL02a}.
The object pursued with this line of research is the development of a
theoretical understanding of possible forms of sequential programs,
starting from the simplest form.
The view is taken that sequential programs in the simplest form are
sequences of instructions.
Program algebra, an algebra of programs in which programs are looked
upon as sequences of instructions, is taken for the basis of the
development aimed at.

The approach to define the meaning of programs followed in this line of
research is called projection semantics.
It explains the meaning of programs in terms of known programs instead
of more or less sophisticated mathematical objects that represent
behaviours of programs under execution.
The main advantage of projection semantics is that it does not require a
lot of mathematical background.
Over and above that, the view is taken that the behaviours of sequential
programs under execution are threads as considered in basic thread
algebra~\cite{BL02a}.%
\footnote
{In~\cite{BL02a}, basic thread algebra is introduced under the name
basic polarized process algebra.
 Prompted by the development of thread algebra~\cite{BM04c}, which is a
 design on top of it, basic polarized process algebra has been renamed
 to basic thread algebra.
}
Therefore, the meaning of the programs considered in program algebra is
explained in terms of threads.
The experience gained so far leads us to believe that sequential
programs are nothing but linear representations of threads.

Sequential programs in the form of assembly programs up to and including
sequential programs in the form of structured programs are covered
in~\cite{BL02a}.
However, although they are found in existing assembly programming
practice, indirect jump instructions are not considered.
In~\cite{BM07e}, several kinds of indirect jump instructions are
considered, including a kind by which recursive method calls can easily
be explained.
Dynamic instruction instantiation is a programming feature that is not
suggested by existing programming practice.
However, from the viewpoint that sequential programs are nothing but
linear representations of threads, it is a genuine programming feature.
It is a useful programming feature as well, as will be illustrated by
means of an example in the paper.
Therefore, we consider a theoretical understanding of instruction
sequences with dynamically instantiated instructions relevant to
programming.

We believe that interaction with services provided by an execution
environment is inherent in the behaviour of programs under execution.
Intuitively, some service provides for dynamic instruction
instantiation.
In this paper, we define the meaning of programs with dynamically
instantiated instructions in two different ways.
In either case, we give a translation by which each program with
dynamically instantiated instructions is turned into a program without
them that exhibits on execution the same behaviour by interaction with
some service.
In one case, the service concerned provides in effect for the dynamic
instruction instantiation and, in the other case, it is largely achieved
by the translated programs.
We consider it useful to treat both cases because of the considerable
difference in complexity of the two translations.

A thread proceeds by doing steps in a sequential fashion.
A thread may do certain steps only for the sake of having itself
affected by some service.
The interaction between behaviours of programs under execution and some
service referred to above is an interaction with that purpose.
In~\cite{BM04c}, the use mechanism is introduced to allow for such a
kind of interaction between threads and services.
In this paper, we will use a generalization of the use mechanism, called
the action transforming use mechanism, to have behaviours of programs
under execution affected by services.
This generalization is reminiscent of the state operator introduced
in~\cite{BB88}.

A hierarchy of program notations rooted in program algebra is introduced
in~\cite{BL02a}.
In this paper, we embroider on one program notation that belongs to this
hierarchy.
The program notation in question, called \PGLD, is a very simple program
notation which is close to existing assembly languages.
The hierarchy also includes a program notation, called \PGLS, that
supports structured programming by offering a rendering of conditional
and loop constructs instead of (unstructured) jump instructions.
Each \PGLS\ program can be translated into a semantically equivalent
\PGLD\ program by means of the projection semantics of \PGLS\ and some
intermediate program notations.

This paper is organized as follows.
First, we review basic thread algebra, program algebra, and program
notation \PGLD\ (Sections~\ref{sect-BTA}, \ref{sect-PGA},
and~\ref{sect-PGLD}).
Next, we provide a simple classification of services that will be used
in subsequent sections (Section~\ref{sect-class-services}).
After that, we extend basic thread algebra with the action transforming
use mechanism and introduce a state-based approach to describe services
(Sections~\ref{sect-TAtsc} and~\ref{sect-service-descr}).
Then, we give a state-based description of a service that can provide
for dynamic instruction instantiation and use that service to define the
meaning of the programs from a variant of the program notation \PGLD\
with dynamically instantiated instructions
(Sections~\ref{sect-meth-act-trl} and~\ref{sect-PGLDdii}).
Following this, we introduce a concrete notation for basic instructions
that covers dynamically instantiated instructions and use that notation
to illustrate the usefulness of dynamic instruction instantiation
(Section~\ref{sect-concr-pinstr}).
Thereupon, we give a state-based description of a register file service
and use that service to define the meaning of the programs from the
variant of the program notation \PGLD\ with dynamically instantiated
instructions in another way (Sections~\ref{sect-reg}
and~\ref{sect-PGLDdii-alt}).
Finally, we discuss the semantic approaches followed in the preceding
sections and make some concluding remarks (Sections~\ref{sect-disc}
and~\ref{sect-concl}).

\section{Basic Thread Algebra}
\label{sect-BTA}

In this section, we review \BTA\ (Basic Thread Algebra), a form of
process algebra which is concerned with the behaviours that sequential
programs exhibit on execution.
The behaviours concerned are called \emph{threads}.
\BTA\ was first presented in~\cite{BL02a}.
In that paper, as well as several other papers, \BTA\ is called \BPPA\
(Basic Polarized Process Algebra).

In \BTA, it is assumed that there is a fixed but arbitrary finite set of
\emph{basic actions} $\BAct$ with $\Tau \not\in \BAct$.
We write $\BActTau$ for $\BAct \union \set{\Tau}$.
The members of $\BActTau$ are referred to as \emph{actions}.

A thread performs basic actions in a sequential fashion.
The intuition is that each basic action performed by a thread is taken
as a command to be processed by a service provided by the execution
environment of the thread.
The processing of a command may involve a change of state of the service
concerned.
At completion of the processing of the command, the service produces a
reply value and returns that reply value to the thread.
The reply value is either $\True$ or $\False$ and determines how the
thread proceeds.

Although \BTA\ is one-sorted, we make this sort explicit.
The reason for this is that we will extend \BTA\ with an additional sort
in Section~\ref{sect-TAtsc}.

The algebraic theory \BTA\ has one sort: the sort $\Thr$ of
\emph{threads}.
To build terms of sort $\Thr$, \BTA\ has the following constants and
operators:
\begin{iteml}
\item
the \emph{inaction} constant $\const{\DeadEnd}{\Thr}$;
\item
the \emph{termination} constant $\const{\Stop}{\Thr}$;
\item
for each $a \in \BActTau$, the binary \emph{postconditional composition}
operator $\funct{\pcc{\ph}{a}{\ph}}{\Thr \x \Thr}{\Thr}$.
\end{iteml}
Terms of sort $\Thr$ are built as usual (see e.g.~\cite{ST99a,Wir90a}).
Throughout the paper, we assume that there are infinitely many variables
of sort $\Thr$, including $x,y,z$.

We use infix notation for postconditional composition.
We introduce \emph{action prefixing} as an abbreviation: $a \bapf p$,
where $p$ is a term of sort $\Thr$, abbreviates $\pcc{p}{a}{p}$.

The thread denoted by a closed term of the form $\pcc{p}{a}{q}$ will
first perform $a$, and then proceed as the thread denoted by $p$ if the
processing of $a$ leads to the reply $\True$ (called a positive reply)
and proceed as the thread denoted by $q$ if the processing of $a$ leads
to the reply $\False$ (called a negative reply).
The action $\Tau$ plays a special role.
It is a concrete internal action: the processing of $\Tau$ will never
involve a state change and always lead to a positive reply, but
notwithstanding all that its presence matters.
The threads denoted by $\DeadEnd$ and $\Stop$ will become inactive and
terminate, respectively.

\BTA\ has only one axiom.
This axiom is given in Table~\ref{axioms-BTA}.%
\begin{table}[!t]
\caption{Axiom of \BTA}
\label{axioms-BTA}
\begin{eqntbl}
\begin{axcol}
\pcc{x}{\Tau}{y} = \pcc{x}{\Tau}{x}                      & \axiom{T1}
\end{axcol}
\end{eqntbl}
\end{table}
Using the abbreviation introduced above, axiom T1 can be written as
follows: $\pcc{x}{\Tau}{y} = \Tau \bapf x$.

Each closed \BTA\ term of sort $\Thr$ denotes a finite thread, i.e.\ a
thread that will become inactive or terminate after it has performed
finitely many actions.
Infinite threads can be described by guarded recursion specifications.

A \emph{guarded recursive specification} over \BTA\ is a set of
recursion equations $E = \set{X = p_X \where X \in V}$, where $V$ is a
set of variables of sort $\Thr$ and each $p_X$ is a term of the form
$\DeadEnd$, $\Stop$ or $\pcc{p}{a}{q}$ with $p$ and $q$ \BTA\ terms
of sort $\Thr$ that contain only variables from $V$.
We write $\vars(E)$ for the set of all variables that occur on the
left-hand side of an equation in $E$.
We are only interested in models of \BTA\ in which guarded recursive
specifications have unique solutions, such as the projective limit model
of \BTA\ presented in~\cite{BB03a}.
A thread that is the solution of a finite guarded recursive
specification over \BTA\ is called a \emph{finite-state} thread.

We extend \BTA\ with guarded recursion by adding constants for solutions
of guarded recursive specifications and axioms concerning these
additional constants.
For each guarded recursive specification $E$ and each $X \in \vars(E)$,
we add a constant of sort $\Thr$ standing for the unique solution of $E$
for $X$ to the constants of \BTA.
The constant standing for the unique solution of $E$ for $X$ is denoted
by $\rec{X}{E}$.
Moreover, we add the axioms for guarded recursion given in
Table~\ref{axioms-REC} to \BTA,%
\begin{table}[!t]
\caption{Axioms for guarded recursion}
\label{axioms-REC}
\begin{eqntbl}
\begin{saxcol}
\rec{X}{E} = \rec{t_X}{E} & \mif X \!=\! t_X \in E       & \axiom{RDP}
\\
E \Implies X = \rec{X}{E} & \mif X \in \vars(E)          & \axiom{RSP}
\end{saxcol}
\end{eqntbl}
\end{table}
where we write $\rec{t_X}{E}$ for $t_X$ with, for all $Y \in \vars(E)$,
all occurrences of $Y$ in $t_X$ replaced by $\rec{Y}{E}$.
In this table, $X$, $t_X$ and $E$ stand for an arbitrary variable of
sort $\Thr$, an arbitrary \BTA\ term of sort $\Thr$ and an arbitrary
guarded recursive specification over \BTA, respectively.
Side conditions are added to restrict the variables, terms and guarded
recursive specifications for which $X$, $t_X$ and $E$ stand.
RDP stands for recursive definition principle and RSP stands for
recursive specification principle.
The equations $\rec{X}{E} = \rec{t_X}{E}$ for a fixed $E$ express that
the constants $\rec{X}{E}$ make up a solution of $E$.
The conditional equations $E \Implies X = \rec{X}{E}$ express that this
solution is the only one.
It is easily demonstrated that RDP and RSP hold in the model of \BTA\
based on projective sequences outlined in~\cite{BL02a} (cf.\ Theorem~1
from~\cite{BM06a}).

We will use the following abbreviation: $a^\omega$, where
$a \in \BActTau$, abbreviates $\rec{X}{\set{X = a \bapf X}}$.

We will write \BTA+\REC\ for \BTA\ extended with the constants for
solutions of guarded recursive specifications and axioms RDP and RSP.

In~\cite{BM05c}, we show that the threads considered in \BTA+\REC\ can
be viewed as processes that are definable over ACP~\cite{Fok00}.

\section{Program Algebra}
\label{sect-PGA}

In this section, we review \PGA\ (ProGram Algebra), an algebra of
sequential programs based on the idea that sequential programs are in
essence sequences of instructions.
\PGA\ provides a program notation for finite-state threads.

In \PGA, it is assumed that there is a fixed but arbitrary finite set
$\BInstr$ of \emph{basic instructions}.
\PGA\ has the following \emph{primitive instructions}:
\begin{iteml}
\item
for each $a \in \BInstr$, a \emph{plain basic instruction} $a$;
\item
for each $a \in \BInstr$, a \emph{positive test instruction} $\ptst{a}$;
\item
for each $a \in \BInstr$, a \emph{negative test instruction} $\ntst{a}$;
\item
for each $l \in \Nat$, a \emph{forward jump instruction} $\fjmp{l}$;
\item
a \emph{termination instruction} $\halt$.
\end{iteml}
We write $\PInstr$ for the set of all primitive instructions.

The intuition is that the execution of a basic instruction $a$ may
modify a state and produces $\True$ or $\False$ at its completion.
In the case of a positive test instruction $\ptst{a}$, basic instruction
$a$ is executed and execution proceeds with the next primitive
instruction if $\True$ is produced; otherwise, the next primitive
instruction is skipped and execution proceeds with the primitive
instruction following the skipped one.
In the case where $\True$ is produced and there is not at least one
subsequent primitive instruction and in the case where $\False$ is
produced and there are not at least two subsequent primitive
instructions, inaction occurs.
In the case of a negative test instruction $\ntst{a}$, the role of the
value produced is reversed.
In the case of a plain basic instruction $a$, the value produced is
disregarded: execution always proceeds as if $\True$ is produced.
The effect of a forward jump instruction $\fjmp{l}$ is that execution
proceeds with the $l$-th next instruction of the program concerned.
If $l$ equals $0$ or the $l$-th next instruction does not exist, then
$\fjmp{l}$ results in inaction.
The effect of the termination instruction $\halt$ is that execution
terminates.

\PGA\ has the following constants and operators:
\begin{iteml}
\item
for each $u \in \PInstr$, an \emph{instruction} constant $u$\,;
\item
the binary \emph{concatenation} operator $\ph \conc \ph$\,;
\item
the unary \emph{repetition} operator $\ph\rep$\,.
\end{iteml}
Terms are built as usual.
Throughout the paper, we assume that there are infinitely many
variables, including $x,y,z$.

We use infix notation for concatenation and postfix notation for
repetition.

Closed \PGA\ terms are considered to denote programs.
The intuition is that a program is in essence a non-empty, finite or
infinite sequence of primitive instructions.
These sequences are called \emph{single pass instruction sequences}
because \PGA\ has been designed to enable single pass execution of
instruction sequences: each instruction can be dropped after it has been
executed.
Programs are considered to be equal if they represent the same single
pass instruction sequence.
The axioms for instruction sequence equivalence are given in
Table~\ref{axioms-PGA}.%
\begin{table}[!t]
\caption{Axioms of \PGA}
\label{axioms-PGA}
\begin{eqntbl}
\begin{axcol}
(x \conc y) \conc z = x \conc (y \conc z)              & \axiom{PGA1} \\
(x^n)\rep = x\rep                                      & \axiom{PGA2} \\
x\rep \conc y = x\rep                                  & \axiom{PGA3} \\
(x \conc y)\rep = x \conc (y \conc x)\rep              & \axiom{PGA4}
\end{axcol}
\end{eqntbl}
\end{table}
In this table, $n$ stands for an arbitrary natural number greater than
$0$.
For each $n > 0$, the term $x^n$ is defined by induction on $n$ as
follows: $x^1 = x$ and $x^{n+1} = x \conc x^n$.
The \emph{unfolding} equation $x\rep = x \conc x\rep$ is
derivable.
Each closed \PGA\ term is derivably equal to a term in
\emph{canonical form}, i.e.\ a term of the form $P$ or $P \conc Q\rep$,
where $P$ and $Q$ are closed \PGA\ terms that do not contain the
repetition operator.

Each closed \PGA\ term is considered to denote a program of which the
behaviour is a finite-state thread, taking the set $\BInstr$ of basic
instructions for the set $\BAct$ of actions.
The \emph{thread extraction} operation $\extr{\ph}$ assigns a thread to
each program.
The thread extraction operation is defined by the equations given in
Table~\ref{axioms-thread-extr} (for $a \in \BInstr$, $l \in \Nat$ and
$u \in \PInstr$)%
\begin{table}[!t]
\caption{Defining equations for thread extraction operation}
\label{axioms-thread-extr}
\begin{eqntbl}
\begin{eqncol}
\extr{a} = a \bapf \DeadEnd \\
\extr{a \conc x} = a \bapf \extr{x} \\
\extr{\ptst{a}} = a \bapf \DeadEnd \\
\extr{\ptst{a} \conc x} =
\pcc{\extr{x}}{a}{\extr{\fjmp{2} \conc x}} \\
\extr{\ntst{a}} = a \bapf \DeadEnd \\
\extr{\ntst{a} \conc x} =
\pcc{\extr{\fjmp{2} \conc x}}{a}{\extr{x}}
\end{eqncol}
\qquad
\begin{eqncol}
\extr{\fjmp{l}} = \DeadEnd \\
\extr{\fjmp{0} \conc x} = \DeadEnd \\
\extr{\fjmp{1} \conc x} = \extr{x} \\
\extr{\fjmp{l+2} \conc u} = \DeadEnd \\
\extr{\fjmp{l+2} \conc u \conc x} = \extr{\fjmp{l+1} \conc x} \\
\extr{\halt} = \Stop \\
\extr{\halt \conc x} = \Stop
\end{eqncol}
\end{eqntbl}
\end{table}
and the rule given in Table~\ref{rule-thread-extr}.%
\begin{table}[!t]
\caption{Rule for infinite jump chains}
\label{rule-thread-extr}
\begin{eqntbl}
\begin{eqncol}
x \scongr \fjmp{0} \conc y \Implies \extr{x} = \DeadEnd
\end{eqncol}
\end{eqntbl}
\end{table}
This rule is expressed in terms of the \emph{structural congruence}
predicate $\ph \scongr \ph$, which is defined by the formulas given in
Table~\ref{axioms-scongr} (for $n,m,l \in \Nat$ and
$u_1,\ldots,u_n,v_1,\ldots,v_{m+1} \in \PInstr$).%
\begin{table}[!t]
\caption{Defining formulas for structural congruence predicate}
\label{axioms-scongr}
\begin{eqntbl}
\begin{eqncol}
\fjmp{n+1} \conc u_1 \conc \ldots \conc u_n \conc \fjmp{0}
\scongr
\fjmp{0} \conc u_1 \conc \ldots \conc u_n \conc \fjmp{0}
\\
\fjmp{n+1} \conc u_1 \conc \ldots \conc u_n \conc \fjmp{m}
\scongr
\fjmp{m+n+1} \conc u_1 \conc \ldots \conc u_n \conc \fjmp{m}
\\
(\fjmp{n+l+1} \conc u_1 \conc \ldots \conc u_n)\rep \scongr
(\fjmp{l} \conc u_1 \conc \ldots \conc u_n)\rep
\\
\fjmp{m+n+l+2} \conc u_1 \conc \ldots \conc u_n \conc
(v_1 \conc \ldots \conc v_{m+1})\rep \scongr {} \\ \hfill
\fjmp{n+l+1} \conc u_1 \conc \ldots \conc u_n \conc
(v_1 \conc \ldots \conc v_{m+1})\rep
\\
x \scongr x
\\
x_1 \scongr y_1 \land x_2 \scongr y_2 \Implies
x_1 \conc x_2 \scongr y_1 \conc y_2 \land
{x_1}\rep \scongr {y_1}\rep
\end{eqncol}
\end{eqntbl}
\end{table}

The equations given in Table~\ref{axioms-thread-extr} do not cover the
case where there is an infinite chain of forward jumps.
Programs are structural congruent if they are the same after removing
all chains of forward jumps in favour of single jumps.
Because an infinite chain of forward jumps corresponds to $\fjmp{0}$,
the rule from Table~\ref{rule-thread-extr} can be read as follows:
if $x$ starts with an infinite chain of forward jumps, then $\extr{x}$
equals $\DeadEnd$.
It is easy to see that the thread extraction operation assigns the same
thread to structurally congruent programs.
Therefore, the rule from Table~\ref{rule-thread-extr} can be replaced by
the following generalization:
$x \scongr y  \Implies \extr{x} = \extr{y}$.

Let $E$ be a finite guarded recursive specification over \BTA, and let
$P_X$ be a closed \PGA\ term for each $X \in \vars(E)$.
Let $E'$ be the set of equations that results from replacing in $E$ all
occurrences of $X$ by $\extr{P_X}$ for each $X \in \vars(E)$.
If $E'$ can be obtained by applications of axioms PGA1--PGA4, the
defining equations for the thread extraction operation and the rule for
infinite jump chains, then $\extr{P_X}$ is the solution of $E$ for $X$.
Such a finite guarded recursive specification can always be found.
Thus, the behaviour of each closed \PGA\ term is a thread that is
definable by a finite guarded recursive specification over \BTA.
Moreover, each finite guarded recursive specification over \BTA\ can be
translated to a closed \PGA\ term of which the behaviour is the solution
of the finite guarded recursive specification concerned
(see Proposition~2 of~\cite{PZ06a}).

Closed \PGA\ terms are loosely called \PGA\ \emph{programs}.
\PGA\ programs in which the repetition operator does not occur are
called \emph{finite} \PGA\ programs.

\section{The Program Notation \PGLD}
\label{sect-PGLD}

In this section, we review a program notation which is rooted in \PGA.
This program notation, called \PGLD, belongs to a hierarchy of program
notations introduced in~\cite{BL02a}.
\PGLD\ is close to existing assembly languages.
It has absolute jump instructions and no explicit termination
instruction.

In \PGLD, as in \PGA, it is assumed that there is a fixed but arbitrary
finite set of \emph{basic instructions} $\BInstr$.
Again, the intuition is that the execution of a basic instruction $a$
may modify a state and produces $\True$ or $\False$ at its completion.

\PGLD\ has the following primitive instructions:
\begin{itemize}
\item
for each $a \in \BInstr$, a \emph{plain basic instruction} $a$;
\item
for each $a \in \BInstr$, a \emph{positive test instruction} $\ptst{a}$;
\item
for each $a \in \BInstr$, a \emph{negative test instruction} $\ntst{a}$;
\item
for each $l \in \Nat$, a \emph{direct absolute jump instruction}
$\ajmp{l}$.
\end{itemize}
\PGLD\ programs have the form $u_1;\ldots;u_k$, where $u_1,\ldots,u_k$
are primitive instructions of \PGLD.
We write $\TProg_\sPGLD$ for the set of all \PGLD\ programs.

The plain basic instructions, the positive test instructions, and the
negative test instructions are as in \PGA.
The effect of a direct absolute jump instruction $\ajmp{l}$ is that
execution proceeds with the $l$-th instruction of the program concerned.
If $\ajmp{l}$ is itself the $l$-th instruction, then inaction occurs.
If $l$ equals $0$ or $l$ is greater than the length of the program, then
termination occurs.

We define the meaning of \PGLD\ programs by means of a function
$\pgldpga$ from the set of all \PGLD\ programs to the set of all \PGA\
programs.
This function is defined by
\begin{ldispl}
\pgldpga(u_1 \conc \ldots \conc u_k) =
(\phi_1(u_1) \conc \ldots \conc \phi_k(u_k) \conc
 \halt \conc \halt)\rep\;,
\end{ldispl}%
where the auxiliary functions $\phi_j$ from the set of all primitive
instructions of \PGLD\ to the set of all primitive instructions of \PGA\
are defined as follows ($1 \leq j \leq k$):
\begin{ldispl}
\begin{aceqns}
\phi_j(\ajmp{l}) & = & \fjmp{l-j}       & \mif j \leq l \leq k\;, \\
\phi_j(\ajmp{l}) & = & \fjmp{k+2-(j-l)} & \mif 0   <  l   <  j\;, \\
\phi_j(\ajmp{l}) & = & \halt            & \mif l = 0 \lor l > k\;, \\
\phi_j(u)        & = & u
                    & \mif u\; \mathrm{is\;not\;a\;jump\;instruction}\;.
\end{aceqns}
\end{ldispl}%

Let $P$ be a \PGLD\ program.
Then $\pgldpga(P)$ represents the meaning of $P$ as a \PGA\ program.
The intended behaviour of $P$ under execution is the behaviour of
$\pgldpga(P)$ under execution.
That is, the \emph{behaviour} of $P$ under execution, written
$\extr{P}_\sPGLD$, is $\extr{\pgldpga(P)}$.

We use the phrase \emph{projection semantics} to refer to the approach
to semantics followed in this section.
The meaning function $\pgldpga$ is called a \emph{projection}.

\sloppy
In the hierarchy of program notations introduced in~\cite{BL02a},
program notations \PGLA, \PGLB\ and \PGLC\ appear between \PGA\ and
\PGLD.
In~\cite{BL02a}, \PGLD\ programs are translated into \PGLC\ programs by
means of a projection $\pgldpglc$, etc.
Above, $\pgldpga$ is defined such that
$\pgldpga(P) = \pglapga(\pglbpgla(\pglcpglb(\pgldpglc(P))))$ for all
\PGLD\ programs $P$.

\section{A Classification of Services}
\label{sect-class-services}

In this short section, we provide a classification of services.
It is a simplified version of the classification given in~\cite{BM07a}
and will be used in subsequent sections.

A distinction is made between target services and para-target services:
\begin{itemize}
\item
A service is a \emph{target service} if the result of the processing of
commands by the service is partly observable externally.
Reading input data from a keyboard, showing output data on a screen and
writing persistent data in permanent memory are typical examples of
using a target service.
\item
A service is a \emph{para-target service} if the result of the
processing of commands by the service is wholly unobservable externally.
Setting a timer and transferring data by means of a Java pipe are
typical examples of using a para-target service.
\end{itemize}

The overall intuition about threads, target services and para-target
services is that:
\begin{itemize}
\item
a thread is the behaviour exhibited by a sequential program on
execution;
\item
a thread interacts with services provided by the execution environment
in question;
\item
the intentions about the resulting behaviour pertain only to interaction
with target services;
\item
interaction with para-target services takes place only in as far as it
is needed to obtain the intended behaviour in relation to target
services.
\end{itemize}

One of the assumptions made in thread algebra is that para-target
services are deterministic.
The exclusion of non-deterministic para-target services is a
simplification.
We believe however that this simplification is adequate in the cases
that we address: para-target services that keep data for a thread.
Of course, it is inadequate in cases where services such as dice-playing
services are taken into consideration.
Another assumption is that target services are non-deterministic.
The reason for this assumption is that the dependence of target services
on external conditions make it appear to threads that they behave
non-deterministically.

\section{An Action Transforming Use Mechanism}
\label{sect-TAtsc}

A thread may perform certain basic actions only for the sake of having
itself affected by a service.
When processing a basic action performed by a thread, a service affects
that thread in one of the following ways:
(i)~by returning a reply value to the thread at completion of the
processing of the basic action;
(ii)~by turning the processed basic action into another basic action.
In this section, we introduce an action transforming use mechanism,
which allows for para-target services to affect threads in either way.
We will only use the action transforming use mechanism to have program
behaviours affected by a para-target service.
The action transforming use mechanism is a generalization of the version
of the use mechanism introduced in~\cite{BM04c}.%
\footnote
{In later papers, this use mechanism is also called thread-service
 composition.}

It is assumed that there is a fixed but arbitrary finite set of
\emph{foci} $\Foci$ and a fixed but arbitrary finite set of
\emph{methods} $\Meth$.
Each focus plays the role of a name of a service provided by the
execution environment that can be requested to process a command.
Each method plays the role of a command proper.
For the set $\BAct$ of basic actions, we take the set
$\set{f.m \where f \in \Foci, m \in \Meth}$.
Performing a basic action $f.m$ is taken as making a request to the
service named $f$ to process command $m$.

We introduce yet another sort: the sort $\Serv$ of \emph{services}.
However, we will not introduce constants and operators to build terms
of this sort.
We identify para-target services with pairs $\tup{H_1,H_2}$, where
$\funct{H_1}{\neseqof{\Meth}}{\set{\True,\False,\Mless,\Blocked}}$ and
$\funct{H_2}{\neseqof{\Meth}}{\BActTau}$,
satisfying the following conditions:%
\footnote
{We write $\seqof{D}$ for the set of all finite sequences with elements
 from set $D$ and we write $\neseqof{D}$ for the set of all non-empty
 finite sequences with elements from set $D$.
 We use the following notation for finite sequences:
 $\emptyseq$ for the empty sequence,
 $\seq{d}$ for the sequence having $d$ as sole element, and
 $\sigma \concat \sigma'$ for the concatenation of finite sequences
 $\sigma$ and $\sigma'$.}
\begin{ldispl}
\Forall{m \in \Meth}{{}} \\ \quad
 {(\Exists{\alpha \in \seqof{\Meth}}
    {H_1(\alpha \concat \seq{m}) = \Mless} \Implies
   \Forall{\alpha' \in \seqof{\Meth}}
    {H_1(\alpha' \concat \seq{m}) \not\in \set{\True,\False}})}\;,
\eqnsep
\Forall{\alpha \in \neseqof{\Meth},m \in \Meth}
 {(H_1(\alpha) = \Blocked \Implies H_1(\alpha \concat \seq{m}) =
   \Blocked)}\;,
\eqnsep
\Forall{\alpha \in \neseqof{\Meth}}
 {(H_1(\alpha) \neq \Mless \Iff H_2(\alpha) = \Tau)}\;.
\end{ldispl}%
$\Mless$ stands for meaningless and $\Blocked$ stands for blocked.
$\Mless$ is used to indicate that a request to process a command is
accepted, but that no reply is produced.
$\Blocked$ is used to indicate that a request to process a command is
rejected.

Let $H$ be a para-target service, and let $H_1$ and $H_2$ be the unique
functions such that $H = \tup{H_1,H_2}$.
Then we write $\rfunc{H}$ and $\afunc{H}$ for $H_1$ and $H_2$,
respectively.
Given a para-target service $H$ and a method $m \in \Meth$,
the \emph{derived service} of $H$ after processing $m$,
written $\derive{m}H$, is defined by
$\rfunc{\derive{m}H}(\alpha) = \rfunc{H}(\seq{m} \concat \alpha)$ and
$\afunc{\derive{m}H}(\alpha) = \afunc{H}(\seq{m} \concat \alpha)$.

A para-target service $H$ can be understood as follows:
\begin{iteml}
\item
if $\rfunc{H}(\seq{m}) \in \set{\True,\False}$, then the request to
process $m$ is accepted by the service, the reply $\rfunc{H}(\seq{m})$
is produced, $m$ is turned into $\Tau$, and the service proceeds as
$\derive{m}H$;
\item
if $\rfunc{H}(\seq{m}) = \Mless$, then the request to process $m$ is
accepted by the service, no reply is produced, $m$ is turned into
$\afunc{H}(\seq{m})$, and the service proceeds as $\derive{m}H$;
\item
if $\rfunc{H}(\seq{m}) = \Blocked$, then the request to process $m$ is
rejected by the service.
\end{iteml}
The three conditions imposed on para-target services can be paraphrased
as follows:
\begin{iteml}
\item
if it is possible that no reply is produced at completion of the
processing of a command, then it is impossible that a positive or
negative reply is produced at completion of the processing of that
command;
\item
after a request to process a command has been rejected, any request to
process a command will be rejected;
\item
a reply is produced at completion of the processing of a command if and
only if the command is turned into $\Tau$.
\end{iteml}

For each $f \in \Foci$, we introduce the binary
\emph{action transforming use} operator
$\funct{\guse{\ph}{f}{\ph}}{\Thr \x \Serv}{\Thr}$.
Intuitively, $\guse{p}{f}{H}$ is the thread that results from processing
all basic actions performed by thread $p$ that are of the form $f.m$ by
the para-target service $H$.
When a basic action of the form $f.m$ performed by thread $p$ is
processed by the para-target service $H$, it is turned into another
action and, if this action is $\Tau$, postconditional composition is
removed in favour of action prefixing on the basis of the reply value
produced.

The axioms for the action transforming use operators are given in
Table~\ref{axioms-mttsc}.%
\begin{table}[!t]
\caption{Axioms for action transforming use operators}
\label{axioms-mttsc}
\begin{eqntbl}
\begin{saxcol}
\guse{\Stop}{f}{H} = \Stop                         & & \axiom{ATU1} \\
\guse{\DeadEnd}{f}{H} = \DeadEnd                   & & \axiom{ATU2} \\
\guse{(\Tau \bapf x)}{f}{H} =
                       \Tau \bapf (\guse{x}{f}{H}) & & \axiom{ATU3} \\
\guse{(\pcc{x}{g.m}{y})}{f}{H} =
\pcc{(\guse{x}{f}{H})}{g.m}{(\guse{y}{f}{H})}
 & \mif f \neq g                                     & \axiom{ATU4} \\
\guse{(\pcc{x}{f.m}{y})}{f}{H} =
\Tau \bapf (\guse{x}{f}{\derive{m}H})
 & \mif \rfunc{H}(\seq{m}) = \True                   & \axiom{ATU5} \\
\guse{(\pcc{x}{f.m}{y})}{f}{H} =
\Tau \bapf (\guse{y}{f}{\derive{m}H})
 & \mif \rfunc{H}(\seq{m}) = \False                  & \axiom{ATU6} \\
\guse{(\pcc{x}{f.m}{y})}{f}{H} =
\\ \hfill
\pcc{(\guse{x}{f}{\derive{m}H})}
    {\afunc{H}(\seq{m})}{(\guse{y}{f}{\derive{m}H})}
 & \mif \rfunc{H}(\seq{m}) = \Mless                  & \axiom{ATU7} \\
\guse{(\pcc{x}{f.m}{y})}{f}{H} = \DeadEnd
 & \mif \rfunc{H}(\seq{m}) = \Blocked                & \axiom{ATU8}
\end{saxcol}
\end{eqntbl}
\end{table}
In this table, $f$ and $g$ stand for arbitrary foci from $\Foci$ and $m$
stands for an arbitrary method from $\Meth$.
Axioms ATU3 and ATU4 express that the action $\Tau$ and basic actions of
the form $g.m$, where $f \neq g$, are not processed.
Axioms ATU5--ATU7 express that a thread is affected by a para-target
service as described above when a basic action of the form $f.m$
performed by the thread is processed by the service.
Axiom ATU8 expresses that inaction occurs when a basic action to be
processed is not accepted.

Let $T$ stand for either \BTA\ or \BTA+\REC.
Then we will write $T$+\ATU\ for $T$, taking the set
$\set{f.m \where f \in \Foci, m \in \Meth}$ for $\BAct$, extended with
the action transforming use operators and the axioms from
Table~\ref{axioms-mttsc}.

The use mechanism introduced in~\cite{BM04c} deals in essence with
para-target services $H$ for which it holds that
$\afunc{H}(\alpha) = \Tau$ for all $\alpha \in \neseqof{\Meth}$.
For these services, the action transforming use mechanism
coincides with the use mechanism from~\cite{BM04c}.

The action $\Tau$ is an internal action whose presence matters.
To conceal its presence in the case where it does not matter after all,
we also introduce the unary \emph{abstraction} operator
$\funct{\abstr}{\Thr}{\Thr}$.

The axioms for the abstraction operator are given in
Table~\ref{axioms-abstr}.%
\begin{table}[!t]
\caption{Axioms for abstraction}
\label{axioms-abstr}
\begin{eqntbl}
\begin{axcol}
\abstr(\Stop) = \Stop                                    & \axiom{TT1} \\
\abstr(\DeadEnd) = \DeadEnd                              & \axiom{TT2} \\
\abstr(\Tau \bapf x) = \abstr(x)                         & \axiom{TT3} \\
\abstr(\pcc{x}{a}{y}) = \pcc{\abstr(x)}{a}{\abstr(y)}    & \axiom{TT4} \\
\abstr(\Tau^\omega) = \DeadEnd                           & \axiom{TT5}
\end{axcol}
\end{eqntbl}
\end{table}
In this table, $a$ stands for an arbitrary basic action from $\BAct$.

Abstraction can for instance be appropriate in cases where $\Tau$ arises
from turning actions of an auxiliary nature into $\Tau$ by means of the
action transforming use mechanism.
In subsequent sections, abstraction will only be used in such cases.
Unlike the use mechanism introduced in~\cite{BM04c}, the use mechanism
introduced in~\cite{BP02a} incorporates abstraction.

Let $T$ stand for either \BTA+\REC\ or \BTA+\REC+\ATU.
Then we will write $T$+\ABSTR\ for $T$ extended with the abstraction
operator and the axioms from Table~\ref{axioms-abstr}.

\section{State-Based Description of Para-Target Services}
\label{sect-service-descr}

In this section, we introduce the state-based approach to describe
families of para-target services that will be used in
Sections~\ref{sect-meth-act-trl} and~\ref{sect-reg}.
This approach is similar to the approach to describe state machines
introduced in~\cite{BP02a}.

In this approach, a family of para-target services is described by
\begin{itemize}
\item
a set of states $S$;
\item
an effect function $\funct{\eff}{\Meth \x S}{S}$;
\item
a yield function
$\funct{\yld}{\Meth \x S}{\set{\True,\False,\Mless,\Blocked}}$;
\item
an action function
$\funct{\act}{\Meth \x S}{\BActTau}$;
\end{itemize}
satisfying the following conditions:
\begin{ldispl}
\Forall{m \in \Meth}
 {(\Exists{s \in S}{\yld(m,s) = \Mless} \Implies
   \Forall{s' \in S}{\yld(m,s') \not\in \set{\True,\False}})}\;,
\eqnsep
\Exists{s \in S}
 {\Forall{m \in \Meth}{{}} \\ \quad
   {(\yld(m,s) = \Blocked \land
     \Forall{s' \in S}
      {(\yld(m,s') = \Blocked \Implies \eff(m,s') = s)})}}\;,
\eqnsep
\Forall{m \in \Meth,s \in S}
 {(\yld(m,s) \neq \Mless \Iff \act(m,s) = \Tau)}\;.
\end{ldispl}%
The set $S$ contains the states in which the services may be, and the
functions $\eff$, $\yld$ and $\act$ give, for each method $m$ and state
$s$, the state, reply and action, respectively, that result from
processing $m$ in state $s$.

We define, for each $s \in S$, a cumulative effect function
$\funct{\ceff_s}{\seqof{\Meth}}{S}$ in terms of $s$ and $\eff$ as follows:
\begin{ldispl}
\ceff_s(\emptyseq) = s\;,
\\
\ceff_s(\alpha \concat \seq{m}) = \eff(m,\ceff_s(\alpha))\;.
\end{ldispl}%
We define, for each $s \in S$, a para-target service $H_s$ in terms of
$\ceff_s$, $\yld$ and $\act$ as follows:
\begin{ldispl}
\rfunc{H_s}(\alpha \concat \seq{m})  = \yld(m,\ceff_s(\alpha))\;,
\\
\afunc{H_s}(\alpha \concat \seq{m}) = \act(m,\ceff_s(\alpha))\;.
\end{ldispl}%
$H_s$ is called the para-target service with \emph{initial state} $s$
described by $S$, $\eff$, $\yld$ and $\act$.
We say that $\set{H_s \where s \in S}$ is the \emph{family of
para-target services} described by $S$, $\eff$, $\yld$ and $\act$.

The conditions that are imposed on $S$, $\eff$, $\yld$ and $\act$ imply
that, for each $s \in S$, $H_s$ is indeed a para-target service.
It is worth mentioning that $\derive{m} H_s = H_{\eff(m,s)}$,
$\rfunc{H_s}(\seq{m}) = \yld(m,s)$, and
$\afunc{H_s}(\seq{m}) = \act(m,s)$.

\section{Method-to-Action Translator Services}
\label{sect-meth-act-trl}

In this section, we give a state-based description of the very simple
family of para-target services that constitute a register-file-dependent
method-to-action translator of which the register file consists of
registers that can contain natural numbers up to some bound.
This method-to-action translator will be used in
Section~\ref{sect-PGLDdii} to describe the behaviour of programs in a
variant of \PGLD\ with dynamically instantiated instructions.

It is assumed that fixed but arbitrary positive numbers
$\maxr,\maxn \in \Nat$ have been given.
$\maxr$ is considered the number of registers in the register file and
$\maxn$ is considered the greatest natural number that can be contained
in the registers of the register file.
The functions from $[1,\maxr]$ to $[0,\maxn]$ are taken for the states
of the register file.
For every function $\funct{s}{[1,\maxr]}{[0,\maxn]}$, $s$ is the state
in which, for each $i \in [1,\maxr]$, the content of register $i$ is
$s(i)$.

It is also assumed that a fixed but arbitrary set
$\PAct \subseteq \Meth$ and a fixed but arbitrary function
$\funct{\theta}{\PAct \x (\mapof{[1,\maxr]}{[0,\maxn]})}{\BAct}$ have
been given.
$\PAct$ is considered the set of methods that are transformable to basic
actions and $\theta$ is regarded to give, for each method $m$ in $\PAct$
and function $\funct{s}{[1,\maxr]}{[0,\maxn]}$, the basic action into
which $m$ is turned in the case where the state of the register file is
$s$.
The methods that belong to $\PAct$ are called \emph{proto-actions}
because they are the methods that are turned into basic actions by the
register-file-dependent method-to-action translator.

The register-file-dependent method-to-action translator services accept
the following methods:
\begin{itemize}
\item
for each $i \in [1,\maxr]$ and $n \in [0,\maxn]$,
a \emph{register set method} $\setr{:}i{:}n$;
\item
each $m \in \PAct$.
\end{itemize}
We write $\Meth_\setr$ for the set
$\set{\setr{:}i{:}n \where i \in [1,\maxr] \land n \in [0,\maxn]}$.
It is assumed that $\Meth_\setr \subseteq \Meth$.

The methods accepted by the method-to-action translator services can be
explained as follows:
\begin{itemize}
\item
$\setr{:}i{:}n$\,:
the content of register $i$ becomes $n$, the reply is $\True$, and
$\setr{:}i{:}n$ is turned into $\Tau$;
\item
$m$, where $m \in \PAct$:
the state of the register file does not change, there is no reply, and
$m$ is turned into $\theta(m,s)$ where $s$ is the state of the register
file.
\end{itemize}

Let $\RFS$ be the set of all functions
$\funct{s}{[1,\maxr]}{[0,\maxn]}$.
Take $\undef$ such that $\undef \notin \RFS$.
Let $s \in \RFS \union \set{\undef}$.
Then we write $\RFDT_s$ for the para-target service with initial state
$s$ described by $S = \RFS \union \set{\undef}$ and the functions
$\eff$, $\yld$, and $\act$ defined as follows
($i \in [1,\maxr]$, $n \in [0,\maxn]$, and $\rho \in \RFS$):%
\footnote
{We use the following notation for functions:
 $f \owr g$ for the function $h$ with $\dom(h) = \dom(f) \union \dom(g)$
 such that for all $d \in \dom(h)$, $h(d) = f(d)$ if $d \not\in \dom(g)$
 and $h(d) = g(d)$ otherwise; and
 $\maplet{d}{r}$ for the function $f$ with $\dom(f) = \set{d}$ such that
 $f(d) = r$.}
\begin{ldispl}
\begin{gceqns}
\multicolumn{2}{@{}l@{}}
{\eff(\setr{:}i{:}n,\rho) = \rho \owr \maplet{i}{n}\;,\footnotemark}
\\
\eff(m,\rho)   = \rho   & \mif m   \in   \PAct\;,
\\
\eff(m,\rho)   = \undef & \mif m \not\in \Meth_\setr \union \PAct\;,
\\
\eff(m,\undef) = \undef\;,
\eqnsep
\yld(\setr{:}i{:}n,\rho) = \True\;,
\\
\yld(m,\rho)   = \Mless   & \mif m   \in   \PAct\;,
\\
\yld(m,\rho)   = \Blocked & \mif m \not\in \Meth_\setr \union \PAct\;,
\\
\yld(m,\undef) = \Blocked\;,
\eqnsep
\act(m,\rho)   = \theta(m,\rho) & \mif m   \in   \PAct\;,
\\
\act(m,\rho)   = \Tau           & \mif m \not\in \PAct\;,
\\
\act(m,\undef) = \Tau\;.
\end{gceqns}
\end{ldispl}%
We write $\RFDT_\mathrm{init}$ for
$\RFDT_{\maplet{1}{0} \owr \ldots \owr \maplet{I}{0}}$.

The special state $\undef$ added above to the proper states of the
register file is a state in which any request to process a method is
rejected.
The existence of such a blocking state is required to guarantee that
$S$, $\eff$, $\yld$ and $\act$ describe a para-target service.

The following proposition states rigorously that the methods that belong
to $\PAct$ are exactly the methods that are turned into basic actions.
\begin{proposition}
For all $\funct{s}{[1,\maxr]}{[0,\maxn]}$:
\begin{ldispl}
\PAct =
 \set{m \in \Meth \where
      \Exists{\alpha \in \seqof{\Meth}}
       {\afunc{\RFDT_s}(\alpha \concat \seq{m}) \in \BAct}}\;.
\end{ldispl}%
\end{proposition}
\begin{proof}
This follows immediately from the definition of the
register-file-dependent method-to-action translator services.
\end{proof}

\section{\PGLD\ with Dynamically Instantiated Instructions}
\label{sect-PGLDdii}

In this section, we introduce a variant of \PGLD\ with dynamically
instantiated instructions.
This variant is called \PGLDdii.
In Section~\ref{sect-concr-pinstr}, the usefulness of dynamic
instruction instantiation will be illustrated by means of an example.

In \PGLDdii, it is assumed that there is a fixed but arbitrary finite set
of \emph{foci} $\Foci$ with $\rfdt \in \Foci$ and a fixed but arbitrary
finite set of \emph{methods} $\Meth$.
Moreover, we adopt the assumptions made about register-file-dependent
method-to-action translator services in Section~\ref{sect-meth-act-trl}.
The set $\set{f.m \where f \in \Foci, m \in \Meth \diff \PAct}$ is taken
as the set $\BInstr$ of basic instructions.
In the setting of \PGLDdii, we use the term \emph{proto-instruction}
instead of proto-action and write $\BPInstr$ instead of $\PAct$.
A proto-instruction is what becomes a basic instruction by dynamic
instantiation.

\PGLDdii\ has the following primitive instructions in addition to the
primitive instructions of \PGLD:
\begin{iteml}
\item
for each $e \in \BPInstr$, a \emph{plain basic proto-instruction} $e$;
\item
for each $e \in \BPInstr$, a \emph{positive test proto-instruction}
$\ptst{e}$;
\item
for each $e \in \BPInstr$, a \emph{negative test proto-instruction}
$\ntst{e}$.
\end{iteml}
\PGLDdii\ programs have the form $u_1 \conc \ldots \conc u_k$, where
$u_1,\ldots,u_k$ are primitive instructions of \PGLDdii.

The effect of a plain basic proto-instruction $e$ is the same as the
effect of the plain basic instruction $\theta(e,s)$, where $s$ is the
state of the register file involved in the instantiation of
proto-instructions.
The effect of a positive or negative test proto-instruction is similar.

Recall that the content of register $i$ can be set to $n$ by means of
the basic instruction $\rfdt.\setr{:}i{:}n$.
Initially, its content is $0$.

We define the meaning of \PGLDdii\ programs by means of a function
$\pglddiipgld$ from the set of all \PGLDdii\ programs to the set of all
\PGLD\ programs.
This function is defined by
\begin{ldispl}
\pglddiipgld(u_1 \conc \ldots \conc u_k) =
\psi(u_1) \conc \ldots \conc \psi(u_k)\;,
\end{ldispl}%
where the auxiliary function $\psi$ from the set of all primitive
instructions of \PGLDdii\ to the set of all primitive instructions of
\PGLD\ is defined as follows:
\begin{ldispl}
\begin{aceqns}
\psi(e)        & = & \rfdt.e        & \mif e \in \BPInstr\;, \\
\psi(\ptst{e}) & = & \ptst{\rfdt.e} & \mif e \in \BPInstr\;, \\
\psi(\ntst{e}) & = & \ntst{\rfdt.e} & \mif e \in \BPInstr\;, \\
\psi(u)        & = & u & \mif u\; \textrm{is not a proto-instruction}\;.
\end{aceqns}
\end{ldispl}%
The idea is that each proto-instruction can be replaced by an
instruction in which the proto-instruction is taken for the method.

Let $P$ be a \PGLDdii\ program.
Then $\pglddiipgld(P)$ represents the meaning of $P$ as a \PGLD\
program.
The intended behaviour of $P$ under execution is the behaviour of
$\pglddiipgld(P)$ under execution on interaction with a
register-file-dependent method-to-action translator when abstracted from
$\Tau$.
That is, the \emph{behaviour} of $P$ under execution, written
$\extr{P}_\sPGLDdii$, is
$\abstr
  (\guse{\extr{\pglddiipgld(P)}_\sPGLD}{\rfdt}{\RFDT_\mathrm{init}})$.

\section{Concrete Proto-Instructions}
\label{sect-concr-pinstr}

At a fairly concrete level, basic instructions and proto-instructions
are strings of characters.
In~\cite{BL02a}, a concrete notation for basic instructions is
introduced for the case where each basic instruction consists of a focus
and a method.
Here, we extend that concrete notation to cover proto-instructions.
The resulting concrete notation will be used in examples of the use
of \PGLDdii.

First of all, we distinguish neutral strings and active strings.
A \emph{neutral string} is an empty string or a string of one or more
characters of which the first character is a letter or a colon and each
of the remaining characters is a letter, a digit or a colon.
An \emph{active string} is a string of two or more characters of which
the first character is an asterisk and each of the remaining characters
is a digit.

A \emph{concrete proto-instruction} is a string of the form $f'.m'$,
where $f'$ and $m'$ are non-empty strings of characters in which neutral
strings and active strings alternate, starting with a neutral string of
which the first character is a letter, and at least one active string
occurs.

A \emph{concrete focus} is a neutral string of which the first character
is a letter.
A \emph{concrete method} is either a neutral string of which the first
character is a letter or a concrete proto-instruction.
A \emph{concrete instruction} is a string of the form $f.m$, where $f$
is a concrete focus and $m$ is a concrete method.

The intention is that instantiation of a concrete proto-instruction
amounts to simultaneously replacing all active strings occurring in it
by strings according to some assignment of strings to active strings.
The assignment concerned must be such that concrete proto-instructions
are turned into concrete instructions.

To accomplish the assignments of strings to active strings, all
active strings of interest must be of the form ${*}\delta$, where
$\delta$ is the decimal representation of some $i \in [1,\maxr]$.
Moreover, an encoding of the assignable strings by numbers in
$[0,\maxn]$ must be given.
Then each state of the register file being involved in \PGLDdii\ induces
an assignment as follows: for each active string of interest, say
${*}\delta$, the string assigned to it is the one that is encoded by the
content of the register with the number of which $\delta$ is the decimal
representation.

The concrete notation for basic instructions introduced above is
sufficiently expressive for all applications that we have in mind.
The assignable strings are in many cases binary or decimal
representations of numbers in the interval $[0,\maxn]$.
In such cases, it is most natural to encode the representations simply
by the numbers that they represent.

\begin{example}
\label{example-password-1}
Consider a program that on execution reads digit by digit the binary
representation of a password and then performs an action to have the
password checked by some para-target service.
The binary representation of a password is a character sequence of a
fixed length, say $n$, of which all characters are among the binary
digits $0$ and $1$.
The program reads in the binary digits which make up the binary
representation of the password by performing actions that are processed
by a target service.
Suppose that the service used for reading in binary digits only accepts
methods of the form $\getb$ and returns the reply $\False$ if the next
binary digit is $0$ and $\True$ if the next binary digit is $1$.
Moreover, suppose that the service used for checking passwords only
accepts methods of the form $\chk{:}\mathit{pw}$, where $\mathit{pw}$ is
the binary representation of a password.
The focus $\stdin$ is used below as a name of the former service and
the focus $\passw$ is used below as a name of the latter service.

In \PGLDdii, where proto-instructions are available, the program has to
distinguish among only $2 \mul n$ cases.
In \PGLD, where no proto-instructions are available, the program has
to distinguish among $2^n$ cases.

Take $\maxr = n$ and $\maxn = 1$.
Consider the case where $n = 3$.
In \PGLDdii, the initial part of the program looks as follows:
\begin{ldispl}
\ptst{\stdin.\getb} \conc \ajmp{5} \conc
\rfdt.\setr{:}1{:}0 \conc \ajmp{6} \conc \rfdt.\setr{:}1{:}1 \conc {} \\
\ptst{\stdin.\getb} \conc \ajmp{10} \conc
\rfdt.\setr{:}2{:}0 \conc \ajmp{11} \conc \rfdt.\setr{:}2{:}1 \conc {} \\
\ptst{\stdin.\getb} \conc \ajmp{15} \conc
\rfdt.\setr{:}3{:}0 \conc \ajmp{16} \conc \rfdt.\setr{:}3{:}1 \conc {} \\
\ptst{\passw.\chk{:}{*}1{:}{*}2{:}{*}3} \conc \ldots
\end{ldispl}%
In \PGLD, the initial part of the program looks as follows:
\begin{ldispl}
\ptst{\stdin.\getb} \conc \ajmp{7} \conc \ajmp{4} \conc {} \\
\ptst{\stdin.\getb} \conc \ajmp{13} \conc \ajmp{10} \conc
\ptst{\stdin.\getb} \conc \ajmp{19} \conc \ajmp{16} \conc {} \\
\ptst{\stdin.\getb} \conc \ajmp{23} \conc \ajmp{22} \conc
\ptst{\stdin.\getb} \conc \ajmp{25} \conc \ajmp{24} \conc {} \\
\ptst{\stdin.\getb} \conc \ajmp{27} \conc \ajmp{26} \conc
\ptst{\stdin.\getb} \conc \ajmp{29} \conc \ajmp{28} \conc {} \\
\ptst{\passw.\chk{:}000} \conc \ajmp{44} \conc \ajmp{45} \conc
\ptst{\passw.\chk{:}001} \conc \ajmp{44} \conc \ajmp{45} \conc {} \\
\ptst{\passw.\chk{:}010} \conc \ajmp{44} \conc \ajmp{45} \conc
\ptst{\passw.\chk{:}011} \conc \ajmp{44} \conc \ajmp{45} \conc {} \\
\ptst{\passw.\chk{:}100} \conc \ajmp{44} \conc \ajmp{45} \conc
\ptst{\passw.\chk{:}101} \conc \ajmp{44} \conc \ajmp{45} \conc {} \\
\ptst{\passw.\chk{:}110} \conc \ajmp{44} \conc \ajmp{45} \conc
\ptst{\passw.\chk{:}111} \conc \ldots
\end{ldispl}%
These programs take $16$ and $43$ instructions, respectively, up to and
including the password-check \mbox{(proto-)}instructions.
In general, we have that:
\begin{iteml}
\item
In \PGLDdii, the program takes $5 \mul n + 1$ instructions up to and
including the password-check proto-instruction;
\item
In \PGLD, the program takes $6 \mul (2^n - 1) + 1$ instructions up to
and including the last password-check instruction.
\end{iteml}
\end{example}

\section{Register File Services}
\label{sect-reg}

In this section, we give a state-based description of the very simple
family of para-target services that constitute a register file
consisting of registers that can contain natural numbers up to some
bound.
This register file will be used in Section~\ref{sect-PGLDdii-alt} to
describe the behaviour of programs in \PGLDdii.

As in Section~\ref{sect-meth-act-trl}, it is assumed that fixed but
arbitrary positive numbers $\maxr,\maxn \in \Nat$ have been given.
$\maxr$ is considered the number of registers in the register file and
$\maxn$ is considered the greatest natural number that can be contained
in the registers of the register file.

The register file services accept the following methods:
\begin{itemize}
\item
for each $i \in [1,\maxr]$ and $n \in [0,\maxn]$,
a \emph{register set method} $\setr{:}i{:}n$;
\item
for each $i \in [1,\maxr]$ and $n \in [0,\maxn]$,
a \emph{register test method} $\eqr{:}i{:}n$.
\end{itemize}
We write $\Meth_\rf$ for the set
$\set{\setr{:}i{:}n,\eqr{:}i{:}n \where
      i \in [1,\maxr] \land n \in [0,\maxn]}$.
It is assumed that $\Meth_\rf \subseteq \Meth$.

The methods accepted by register services can be explained as follows:
\begin{itemize}
\item
$\setr{:}i{:}n$\,:
the content of register $i$ becomes $n$, the reply is $\True$, and
$\setr{:}i{:}n$ is turned into $\Tau$;
\item
$\eqr{:}i{:}n$\,:
the content of the register does not change, the reply is $\True$ if the
content of register $i$ equals $n$ and $\False$ otherwise, and
$\eqr{:}i{:}n$ is turned into $\Tau$.
\end{itemize}

Let $\RFS$ be the set of all functions
$\funct{s}{[1,\maxr]}{[0,\maxn]}$.
Take $\undef$ such that $\undef \notin \RFS$.
Let $s \in \RFS \union \set{\undef}$.
Then we write $\RF_s$ for the para-target service with initial state
$s$ described by $S = \RFS \union \set{\undef}$ and the functions
$\eff$, $\yld$, and $\act$ defined as follows
($i \in [1,\maxr]$, $n \in [0,\maxn]$, and $\rho \in \RFS$):
\begin{ldispl}
\begin{gceqns}
\eff(\setr{:}i{:}n,\rho) = \rho \owr \maplet{i}{n}\;,
\\
\eff(\eqr{:}i{:}n,\rho)  = \rho\;,
\\
\eff(m,\rho)             = \undef   & \mif m \not\in \Meth_\rf\;,
\\
\eff(m,\undef)           = \undef\;,
\eqnsep
\yld(\setr{:}i{:}n,\rho) = \True    \;,
\\
\yld(\eqr{:}i{:}n,\rho)  = \True    & \mif \rho(i) = n\;,
\\
\yld(\eqr{:}i{:}n,\rho)  = \False   & \mif \rho(i) \neq n\;,
\\
\yld(m,\rho)             = \Blocked & \mif m \not\in \Meth_\rf\;,
\\
\yld(m,\undef)           = \Blocked\;,
\eqnsep
\act(m,\rho)  = \Tau\;,
\\
\act(m,\undef) = \Tau\;.
\end{gceqns}
\end{ldispl}%
We write $\RF_\mathrm{init}$ for
$\RF_{\maplet{1}{0} \owr \ldots \owr \maplet{I}{0}}$.

\section{An Alternative Semantics for \PGLDdii}
\label{sect-PGLDdii-alt}

In this section, we discuss an alternative semantics for \PGLDdii.

Unlike the meaning of \PGLDdii\ programs that we defined in
Section~\ref{sect-PGLDdii}, we define the alternative meaning of
\PGLDdii\ programs only for the case where $\maxr = 1$.
The generalization of the definition to arbitrary $\maxr$ is obvious,
but leads to a definition that is hard to read.

The alternative meaning of \PGLDdii\ programs is given by a function
$\pglddiipgld'$ from the set of all \PGLDdii\ programs to the set of all
\PGLD\ programs.
For the case where $\maxr = 1$, this function is defined by
\begin{ldispl}
\pglddiipgld'(u_1 \conc \ldots \conc u_k) =
\psi'_1(u_1) \conc \ldots \conc \psi'_k(u_k)\;,
\end{ldispl}%
where the auxiliary functions $\psi'_j$ from the set of all primitive
instructions of \PGLDdii\ to the set of all \PGLD\ programs are defined
as follows ($1 \leq j \leq k$):
\begin{ldispl}
\begin{aeqns}
\psi'_j(e) & = &
\ptst{\rf.\eqr{:}1{:}0} \conc \ajmp{l''_{j,0}} \conc {}
\\ & & \quad \vdots
\\ & &
\ptst{\rf.\eqr{:}1{:}\maxn{-}1} \conc \ajmp{l''_{j,\maxn{-}1}} \conc
\ajmp{l''_{j,\maxn}} \conc {}
\\ & &
\theta(e,0) \conc \ajmp{l'_{j+1}} \conc \ajmp{l'_{j+2}} \conc {}
\\ & & \quad \vdots
\\ & &
\theta(e,\maxn{-}1) \conc \ajmp{l'_{j+1}} \conc \ajmp{l'_{j+2}} \conc
\theta(e,\maxn)\;, \\
\psi'_j(\ptst{e}) & = &
\ptst{\rf.\eqr{:}1{:}0} \conc \ajmp{l''_{j,0}} \conc {}
\\ & & \quad \vdots
\\ & &
\ptst{\rf.\eqr{:}1{:}\maxn{-}1} \conc \ajmp{l''_{j,\maxn{-}1}} \conc
\ajmp{l''_{j,\maxn}} \conc {}
\\ & &
\ptst{\theta(e,0)} \conc \ajmp{l'_{j+1}} \conc \ajmp{l'_{j+2}} \conc {}
\\ & & \quad \vdots
\\ & &
\ptst{\theta(e,\maxn{-}1)} \conc \ajmp{l'_{j+1}} \conc \ajmp{l'_{j+2}}
 \conc \ptst{\theta(e,\maxn)}\;, \\
\psi'_j(\ntst{e}) & = &
\ptst{\rf.\eqr{:}1{:}0} \conc \ajmp{l''_{j,0}} \conc {}
\\ & & \quad \vdots
\\ & &
\ptst{\rf.\eqr{:}1{:}\maxn{-}1} \conc \ajmp{l''_{j,\maxn{-}1}} \conc
\ajmp{l''_{j,\maxn}} \conc {}
\\ & &
\ntst{\theta(e,0)} \conc \ajmp{l'_{j+1}} \conc \ajmp{l'_{j+2}} \conc {}
\\ & & \quad \vdots
\\ & &
\ntst{\theta(e,\maxn{-}1)} \conc \ajmp{l'_{j+1}} \conc \ajmp{l'_{j+2}}
 \conc \ntst{\theta(e,\maxn)}\;, \\
\psi'_j(\rfdt.m) & = & \rf.m\;, \\
\psi'_j(\ptst{\rfdt.m}) & = & \ptst{\rf.m}\;, \\
\psi'_j(\ntst{\rfdt.m}) & = & \ntst{\rf.m}\;, \\
\psi'_j(\ajmp{l}) & = & \ajmp{l'_l}\;, \\
\psi'_j(u) & = & u
\qquad\quad
 \mif u\; \textrm{is not a proto-instruction, jump instruction or}
\\ & & \phantom{u \qquad\quad \mif}
\textrm{a plain basic or test instruction with focus}\; \rfdt\;,
\end{aeqns}
\end{ldispl}%
and for each $j \in [1,k]$ and $h \in [0,\maxn]$:
\begin{ldispl}
\begin{aeqns}
l'_j      & = & j + (5 \mul \maxn + 1) \mul n_j\;, \\
l''_{j,h} & = & l'_j + 2 \mul \maxn + 3 \mul h + 1\;,
\end{aeqns}
\end{ldispl}%
and $n_j$ is the number of proto-instructions preceding position $j$.

The idea is that each proto-instruction can be replaced by an
instruction sequence of which the execution leads to the execution of
the intended instruction after the content of the register has been
found by a linear search.
Because the length of the replacing instruction sequence is greater than
$1$, the direct absolute jump instructions are adjusted so as to
compensate for the introduction of additional instructions.
Obviously, the linear search for the content of the register can be
replaced by a binary search.

Henceforth, we will proceed as if $\pglddiipgld'$ has been defined for
arbitrary~$\maxr$.

Let $P$ be a \PGLDdii\ program.
Then $\pglddiipgld'(P)$ represents an alternative meaning of $P$ as a
\PGLD\ program.
The alternative behaviour of $P$ under execution is the behaviour of
$\pglddiipgld'(P)$ under execution on interaction with a register file
when abstracted from $\Tau$.
That is, the \emph{alternative behaviour} of $P$ under execution,
written $\extr{P}'_\sPGLDdii$, is
$\abstr(\guse{\extr{\pglddiipgld'(P)}_\sPGLD}{\rf}{\RF_\mathrm{init}})$.

\begin{example}
Consider the \PGLDdii\ program from Example~\ref{example-password-1}.
The initial part of the \PGLD\ program that results from its translation
by means of $\pglddiipgld$ looks as follows:
\begin{ldispl}
\ptst{\stdin.\getb} \conc \ajmp{5} \conc
\rfdt.\setr{:}1{:}0 \conc \ajmp{6} \conc \rfdt.\setr{:}1{:}1 \conc {} \\
\ptst{\stdin.\getb} \conc \ajmp{10} \conc
\rfdt.\setr{:}2{:}0 \conc \ajmp{11} \conc \rfdt.\setr{:}2{:}1 \conc {} \\
\ptst{\stdin.\getb} \conc \ajmp{15} \conc
\rfdt.\setr{:}3{:}0 \conc \ajmp{16} \conc \rfdt.\setr{:}3{:}1 \conc {} \\
\ptst{\rfdt.\passw.\chk{:}{*}1{:}{*}2{:}{*}3} \conc \ldots
\end{ldispl}%
The initial part of the \PGLD\ program that results from its translation
by means of $\pglddiipgld'$ looks as follows:
\begin{ldispl}
\ptst{\stdin.\getb} \conc \ajmp{5} \conc
\rf.\setr{:}1{:}0 \conc \ajmp{6} \conc \rf.\setr{:}1{:}1 \conc {} \\
\ptst{\stdin.\getb} \conc \ajmp{10} \conc
\rf.\setr{:}2{:}0 \conc \ajmp{11} \conc \rf.\setr{:}2{:}1 \conc {} \\
\ptst{\stdin.\getb} \conc \ajmp{15} \conc
\rf.\setr{:}3{:}0 \conc \ajmp{16} \conc \rf.\setr{:}3{:}1 \conc {} \\
\ptst{\rf.\eqr{:}1{:}0} \conc
\ajmp{19} \conc \ajmp{22} \conc {} \\
\ptst{\rf.\eqr{:}2{:}0} \conc
\ajmp{25} \conc \ajmp{28} \conc {} \\
\ptst{\rf.\eqr{:}2{:}0} \conc
\ajmp{31} \conc \ajmp{34} \conc {} \\
\ptst{\rf.\eqr{:}3{:}0} \conc
\ajmp{37} \conc \ajmp{40} \conc {} \\
\ptst{\rf.\eqr{:}3{:}0} \conc
\ajmp{43} \conc \ajmp{46} \conc {} \\
\ptst{\rf.\eqr{:}3{:}0} \conc
\ajmp{49} \conc \ajmp{52} \conc {} \\
\ptst{\rf.\eqr{:}3{:}0} \conc
\ajmp{55} \conc \ajmp{58} \conc {} \\
\ptst{\passw.\chk{:}000} \conc
\ajmp{59} \conc \ajmp{60} \conc {} \\
\ptst{\passw.\chk{:}001} \conc
\ajmp{59} \conc \ajmp{60} \conc {} \\
\ptst{\passw.\chk{:}010} \conc
\ajmp{59} \conc \ajmp{60} \conc {} \\
\ptst{\passw.\chk{:}011} \conc
\ajmp{59} \conc \ajmp{60} \conc {} \\
\ptst{\passw.\chk{:}100} \conc
\ajmp{59} \conc \ajmp{60} \conc {} \\
\ptst{\passw.\chk{:}101} \conc
\ajmp{59} \conc \ajmp{60} \conc {} \\
\ptst{\passw.\chk{:}110} \conc
\ajmp{59} \conc \ajmp{60} \conc {} \\
\ptst{\passw.\chk{:}111} \conc {} \ldots
\end{ldispl}%
These \PGLD\ programs take $16$ and $58$ instructions, respectively, up
to and including the password-check instructions.

Let $b_1$, $b_2$ and $b_3$ be either $0$ or $1$.
Suppose that the three bits read in at the beginning of the execution of
these programs are $b_1$, $b_2$ and $b_3$, in that order.
In the case of the former program, it is easy to check that the
instruction $\ptst{\rfdt.\passw.\chk{:}{*}1{:}{*}2{:}{*}3}$ will be
executed while the contents of registers $1$, $2$ and $3$ are $b_1$,
$b_2$ and $b_3$, respectively.
In the case of the latter program, it is easy to check that the
instruction $\ptst{\passw.\chk{:}b_1b_2b_3}$ will be executed after
execution of some test and jump instructions.
This strongly suggests that the programs are ``behaviourally
equivalent''.
\end{example}

The following theorem states rigorously that, for any \PGLDdii\ program,
the behaviour under execution coincides with the alternative behaviour
under execution.

\begin{theorem}
\label{theorem-behaviour}
For all \PGLDdii\ programs $P$,
$\extr{P}_\sPGLDdii = \extr{P}'_\sPGLDdii$.
\end{theorem}
\begin{proof}
Strictly speaking, we prove this theorem in the algebraic theory
obtained by:
(i)~combining \PGA\ with \BTA+\REC+\ATU+\ABSTR, resulting in a theory
with three sorts: a sort $\Prog$ of programs, a sort $\Thr$ of threads,
and a sort $\Serv$ of services;
(ii)~extending the result by taking $\extr{\ph}$ for an additional
operator from sort $\Prog$ to sort $\Thr$ and taking the semantic
equations and rule defining thread extraction for additional axioms.
We write $\TThr$ for the set of all closed terms of sort $\Thr$ from the
language of the resulting theory.

In the proof, we make use of an auxiliary function
$\funct{\extr{\ph,\ph}}{\Nat \x \TProg_\sPGLD}{\TThr}$
which gives, for each natural number $i$ and \PGLD\ program
$u_1 \conc \ldots \conc u_k$, a closed term of sort $\Thr$ that denotes
the behaviour of $u_1 \conc \ldots \conc u_k$ when executed from
position $i$ if $1 \leq i \leq k$ and $\Stop$ otherwise.
This function is defined as follows:
\begin{ldispl}
\begin{aceqns}
\extr{i,u_1 \conc \ldots \conc u_k} & = &
\multicolumn{2}{@{}l@{}}
{\extr{\phi_i(u_i) \conc \ldots \conc \phi_k(u_k) \conc
       \halt \conc \halt \conc
       (\phi_1(u_1) \conc \ldots \conc \phi_k(u_k) \conc
        \halt \conc \halt)\rep}
} \\ & &
& \hspace*{17.25em} \mif 1 \leq i \leq k\;,
\\
\extr{i,u_1 \conc \ldots \conc u_k} & =  & \Stop
& \hspace*{17.25em} \mif \Not 1 \leq i \leq k\;
\end{aceqns}
\end{ldispl}%
(where $\phi_j$ is as in the definition of $\pgldpga$).
It follows easily from the definition of $\extr{\ph,\ph}$ and the axioms
of \PGA\ that if $1 \leq i \leq k$:
\begin{ldispl}
\begin{aceqns}
\extr{i,u_1 \conc \ldots \conc u_k} & = &
a \bapf \extr{i+1,u_1 \conc \ldots \conc u_k}
& \mif u_i = a\;, \\
\extr{i,u_1 \conc \ldots \conc u_k} & = &
\pcc{\extr{i+1,u_1 \conc \ldots \conc u_k}}{a}
    {\extr{i+2,u_1 \conc \ldots \conc u_k}}
& \mif u_i = \ptst{a}\;, \\
\extr{i,u_1 \conc \ldots \conc u_k} & = &
\pcc{\extr{i+2,u_1 \conc \ldots \conc u_k}}{a}
    {\extr{i+1,u_1 \conc \ldots \conc u_k}}
& \mif u_i = \ntst{a}\;, \\
\extr{i,u_1 \conc \ldots \conc u_k} & = &
\extr{l,u_1 \conc \ldots \conc u_k}
& \mif u_i = \ajmp{l}\;.
\end{aceqns}
\end{ldispl}%

Let $v_1,\ldots,v_k$ be primitive instructions of \PGLDdii, let
\begin{ldispl}
T =
\set{\abstr
      (\guse{\extr{i,\psi(v_1) \conc \ldots \conc \psi(v_k)}}
            {\rfdt}{\RFDT_s})
     \where i \in [1,k] \land \funct{s}{[1,\maxr]}{[0,\maxn]}}\;, \\
T' =
\set{\abstr
      (\guse{\extr{l'_i,\psi'_1(v_1) \conc \ldots \conc \psi'_k(v_k)}}
            {\rf}{\RF_s})
     \where i \in [1,k] \land \funct{s}{[1,\maxr]}{[0,\maxn]}}\;
\end{ldispl}%
(where $\psi$, $\psi'_j$, $l'_i$ are as in the definitions of
$\pglddiipgld$ and $\pglddiipgld'$),
and let $\funct{\beta}{T}{T'}$ be the bijection defined by
\begin{ldispl}
\beta(\abstr
       (\guse{\extr{i,\psi(v_1) \conc \ldots \conc \psi(v_k)}}
             {\rfdt}{\RFDT_s}))
\\ \quad {} =
\abstr(\guse{\extr{l'_i,\psi'_1(v_1) \conc \ldots \conc \psi'_k(v_k)}}
            {\rf}{\RF_s})\;.
\end{ldispl}%
For each $p' \in \TThr$, write $\beta^*(p')$ for $p'$ with, for all
$p \in T$, all occurrences of $p$ in $p'$ replaced by $\beta(p)$.
Then, using the equations concerning the auxiliary function
$\extr{\ph,\ph}$ given above, it is straightforward to prove that there
exists a set~$E$ consisting of one derivable equation $p = p'$ for each
$p \in T$ such that, for all equations $p = p'$ in $E$:
\begin{iteml}
\item
the equation $\beta(p) = \beta^*(p')$ is derivable;
\item
$p' \in T$ only if $p'$ can be rewritten to a $p'' \not\in T$ using the
equations in $E$ from left to right.
\end{iteml}
Because
$\extr{\psi(v_1) \conc \ldots \conc \psi(v_k)} =
 \extr{1,\psi(v_1) \conc \ldots \conc \psi(v_k)}$ and
$\extr{\psi'_1(v_1) \conc \ldots \conc \psi'_k(v_k)} =
 \extr{l'_1,\psi'_1(v_1) \conc \ldots \conc \psi'_k(v_k)}$,
this means that
$\extr{v_1 \conc \ldots \conc v_k}_\sPGLDdii$ and
$\extr{v_1 \conc \ldots \conc v_k}'_\sPGLDdii$ are solutions of the same
guarded recursive specification.
Because guarded recursive specifications have unique solutions, it
follows immediately that
$\extr{v_1 \conc \ldots \conc v_k}_\sPGLDdii =
 \extr{v_1 \conc \ldots \conc v_k}'_\sPGLDdii$.
\end{proof}

\section{Discussion of Semantic Approaches}
\label{sect-disc}

In Sections~\ref{sect-PGLDdii} and~\ref{sect-PGLDdii-alt}, the meaning
of \PGLDdii\ programs is explained by means of different translations
into \PGLD\ programs.
In both sections, the intended behaviour of a \PGLDdii\ program under
execution is described as the behaviour of the translated program under
execution on interaction with some para-target service.
The translation from Section~\ref{sect-PGLDdii} is extremely simple,
but the translation from Section~\ref{sect-PGLDdii-alt} is fairly
complicated.
The para-target service used in Section~\ref{sect-PGLDdii} to describe
the behaviour of a \PGLDdii\ program and the one used in
Section~\ref{sect-PGLDdii-alt} are equally simple.
However, the former service is far more powerful: it turns a processed
method into a basic action if the method corresponds to a
proto-instruction.
By its power, the translation can be kept simple if that service is
used.
Because of the simpler translation of \PGLDdii\ programs into \PGLD\
programs and the equally simple para-target service used, the approach
followed in Section~\ref{sect-PGLDdii} to define the meaning of
\PGLDdii\ programs is preferable.

A manifestation of the difference in complexity of the translations of
\PGLDdii\ programs from Sections~\ref{sect-PGLDdii}
and~\ref{sect-PGLDdii-alt} is that the former translation replaces each
primitive instruction of \PGLDdii\ by one primitive instruction of
\PGLD, whereas the latter translation gives rise to a combinatorial
explosion.
Recall that $\maxr$ stands for the number of registers involved in the
instantiation of proto-instructions and $\maxn$ stands for the greatest
natural number that can be contained in those registers.
The translation from Section~\ref{sect-PGLDdii-alt} replaces each
primitive instruction of \PGLDdii\ that is not a proto-instruction by
one primitive instruction of \PGLD\ as well, but replaces each
proto-instruction by a sequence of
\begin{ldispl}
(2 \mul \maxn +1) \mul \sum_{i=1}^{\maxr} (\maxn + 1)^{i-1} +
 3 \mul ((\maxn + 1)^\maxr - 1) + 1
\\ \;\; {} =
 (5 \mul \maxn +1) \mul (((\maxn + 1)^\maxr - 1) / \maxn) + 1
\end{ldispl}%
primitive instructions of \PGLD.

If a new programming feature is added to a known program notation such
as \PGLD\ and the starting point of the approach to define the meaning
of the programs from the extended program notation is translation of
those programs into programs from the known program notation, then we
can conceive of several approaches:
\begin{iteml}
\item
give a translation by which each program from the extended program
notation is translated into a program from the known program notation
that exhibits on execution the same behaviour;
\item
give a translation by which each program from the extended program
notation is translated into a program from the known program notation
that exhibits on execution the same behaviour by interaction with a
given para-target service that does not turn any processed method into a
basic action;
\item
give a translation by which each program from the extended program
notation is translated into a program from the known program notation
that exhibits on execution the same behaviour by interaction with a
given para-target service that turns certain processed methods into
basic actions.
\end{iteml}
We consider an approach earlier in this list preferable provided that
the translation concerned does not become too complicated.
In the case where the translation becomes too complicated with all three
approaches, it is desirable to look for another starting point.
This may end up in direct thread extraction, i.e.\ assigning a thread to
each program as this was done for \PGA\ in Section~\ref{sect-PGA}.

In the case of \PGLDdii, it is obvious that the first approach in the
list given above does not work.
However, it is virtually impossible to find out that the third approach
is preferable to the second one without actually producing definitions
of the meaning of \PGLDdii\ programs according to both approaches.

\section{Conclusions}
\label{sect-concl}

We have studied sequential programs that are instruction sequences with
dynamically instantiated instructions.
We have defined the meaning of the programs concerned in two different
ways, which both involve a translation into programs that are
instruction sequences without dynamically instantiated instructions.
In one of the two ways, the translation is very simple and does not lead
to increase in the length of a program or the number of steps needed by
a program.
That way is considered the preferred one.
The preferred way made it necessary for the use mechanism that was
introduced in~\cite{BM04c} to be generalized.
In~\cite{BM08a}, we demonstrate that dynamic instruction instantiation
is a useful programming feature.

In this paper, we have followed the approach of projection semantics,
starting from the perception of a program as an instruction sequence.
This means that programs are considered at a much lower level than usual
in theoretical computer science.
This allows for bringing the interface between software and hardware
better into the picture, which becomes increasingly important to a
growing number of developments related to computer architecture.
The usual approaches to define the meaning of programs, such as
operational semantics, denotational semantics and axiomatic semantics,
are based on the view that the details of program execution should be
abstracted from as much as possible.
This makes comparisons with those approaches virtually impossible.

In~\cite{BM06b}, we have modelled and analysed micro-architectures with
pipelined instruction processing in the setting of program algebra,
basic thread algebra, and Maurer computers~\cite{Mau66a,Mau06a}.
In that work, which we consider a preparatory step in the development of
a formal approach to design new micro-architectures, dynamically
instantiated instructions were not taken into account.
An option for future work is to look at the effect of dynamically
instantiated instructions on pipelined instruction processing.

\bibliographystyle{spmpsci}
\bibliography{TA}

\end{document}